\documentclass[12pt]{article}
\RequirePackage{cite}
\usepackage{times}
\usepackage{fullpage}
\usepackage{ifthen}
\usepackage{graphicx}
\usepackage{enumerate}
\usepackage{amsmath}
\usepackage[UKenglish]{babel}
\usepackage[latin1]{inputenc}
\usepackage[font=small,labelfont=bf]{caption}
\makeatletter
\newcommand{\spacing}[1]{\renewcommand{\baselinestretch}{#1}\large\normalsize}

\renewcommand\@biblabel[1]{#1.}

\def\@cite#1#2{$^{\mbox{\scriptsize #1\if@tempswa , #2\fi}}$}

\def\@maketitle{%
  \newpage\spacing{1}\setlength{\parskip}{12pt}%
    {\Large\bfseries\noindent\sloppy \textsf{\@title} \par}%
    {\noindent\sloppy \@author}%
}

\newenvironment{affiliations}{%
    \setcounter{page}{1}%
    \setlength{\parindent}{0in}%
    \slshape\sloppy%
    \begin{list}{\upshape$^{\arabic{enumi}}$}{%
        \usecounter{enumi}%
        \setlength{\leftmargin}{0in}%
        \setlength{\topsep}{0in}%
        \setlength{\labelsep}{0in}%
        \setlength{\labelwidth}{0in}%
        \setlength{\listparindent}{0in}%
        \setlength{\itemsep}{0ex}%
        \setlength{\parsep}{0in}%
        }
    }{\end{list}\par\vspace{12pt}}

\renewcommand{\section}{\@startsection{section}{1}{0pt}{-6pt}{-1pt}{\textbf}}
\renewcommand{\subsection}{\@startsection{subsection}{2}{0pt}{-0pt}{-0.5em}{\textbf}}

\newenvironment{methods}{%
    \section*{Methods}%
    \setlength{\parskip}{12pt}%
    }{}

\newenvironment{addendum}{%
    \setlength{\parindent}{0in}%
    \small%
    \begin{list}{Acknowledgements}{%
        \setlength{\leftmargin}{0in}%
        \setlength{\listparindent}{0in}%
        \setlength{\labelsep}{0em}%
        \setlength{\labelwidth}{0in}%
        \setlength{\itemsep}{12pt}%
        }
    }
    {\end{list}\normalsize}

\makeatletter

\renewcommand{\subsection}{\@startsection {subsection}{2}{0pt}{-0pt}{-0.5em}{\textbf}*}
\newcommand{\Imodamp}{\left. I_{\mathrm{0}}\right._{\mathrm{inc}}^{\mathrm{mod}}}
\newcommand{\Idcamp}{\left. I_{\mathrm{0}}\right._{\mathrm{inc}}^{\mathrm{unmod}}}
\newcommand{\transition}{58S_{1/2} \rightarrow 59S_{1/2}}

\newcommand{\Pmodamp}{\left. P_{\mathrm{0}}\right._{\mathrm{low}}}
\newcommand{\Pdcamp}{\left. P_{\mathrm{0}}\right._{\mathrm{high}}}
\newcommand{\Iomega}{\left. I_{\mathrm{\Omega}}\right._{\mathrm{sw}}^{\mathrm{mod+unmod}}}
\newcommand{\citeKnuffman}{[1$^*$]}
\newcommand{\citeYounge}{[2$^*$]}
\newcommand{\citeKnuffmanspace}{[1$^*$] }

\makeatother
\title{Forbidden atomic transitions driven by an intensity-modulated laser trap}

\author{Kaitlin~R.~Moore, Sarah~E.~Anderson \& Georg~Raithel}

\begin{document}

\maketitle

\begin{affiliations}
\item[] {\footnotesize Department of Physics and Program in Applied Physics, University of Michigan, Ann Arbor, Michigan 48109, USA}
\end{affiliations}

\spacing{1}
\begin{abstract}
Spectroscopy is an essential tool in understanding and manipulating quantum systems, such as atoms and molecules.  The model describing spectroscopy includes a multipole-field interaction, which leads to established spectroscopic selection rules, and an interaction that is quadratic in the field, which is often neglected.  However, spectroscopy using the quadratic (ponderomotive) interaction promises two significant advantages over spectroscopy using the multipole-field interaction: flexible transition rules and vastly improved spatial addressability of the quantum system.  For the first time, we demonstrate ponderomotive spectroscopy by using optical-lattice-trapped Rydberg atoms, pulsating the lattice light at a microwave frequency, and driving a microwave atomic transition that would otherwise be forbidden by established spectroscopic selection rules.  This new ability to measure frequencies of previously inaccessible transitions makes possible improved determinations of atomic characteristics and constants underlying physics.  In the spatial domain, the resolution of ponderomotive spectroscopy is orders of magnitude better than the transition frequency (and the corresponding diffraction limit) would suggest, promising single-site addressability in a dense particle array for quantum control and computing applications.  Future advances in technology may allow ponderomotive spectroscopy to be extended to ground-state atoms and trapped molecules.

\end{abstract}

\spacing{1.5}
\subsection{Introduction.}
Spectroscopy is a well-established, powerful tool in science for characterizing microscopic systems.  Fields that use this tool range from precision metrology\cite{Ye.2008} (development of frequency standards\cite{Katori.2011} and high-precision sensing, such as gravitometry\cite{Chu.2001}) to trace analysis\cite{Lu.2010} and chemical sensing\cite{Schliesser.2005}.  Characterizing and manipulating particles using light is also the foundation of new fields such as quantum optics\cite{Berman.2011}, computing\cite{Jaksch.2000}, and information processing\cite{Saffman.2010}$^,$\cite{Kimble.2008}.  The interaction between a particle and a light field is described by the particle-field interaction Hamiltonian\cite{Sakurai.1967}, which includes a mulipole-field interaction term and a ponderomotive (quadratic) interaction term.  In the case of direct application of a radiation field to atoms, the dipole-field term typically dominates the total atom-field interaction and leads to the electric-dipole selection rules for atomic transitions.  In contrast, the ponderomotive term dominates the atom-field interaction when there is substantial spatial variation of the field intensity within the volume of the atom and when the intensity is modulated in time at the transition frequency of interest.  In this case, any applicable selection rules are much less restrictive.

To experimentally demonstrate an atomic transition via the ponderomotive interaction for the first time, we have chosen to use Rydberg atoms trapped in an intensity-modulated standing-wave optical lattice.  Rydberg-atom optical lattices are an ideal tool for this demonstration because the atoms' electronic probability distributions can extend over several wells of the optical lattice\cite{Anderson.2011}, and Rydberg-Rydberg transitions are in the microwave regime\cite{Gallagher.1994}, a regime in which light-modulation technology exists.  Using this system, we have demonstrated for the first time a specific case of ponderomotive spectroscopy.  The approach used here may be extended to other particles, including ground-state atoms and trapped molecules, with appropriate advances in technology.

Advantages of ponderomotive spectroscopy over typical dipole-field spectroscopy include flexible transition rules and vastly improved spatial addressability.  First, ponderomotive spectroscopy affords single-step access to energy transitions that are forbidden by established dipole selection rules.  When a particle size and a standing-wave period are comparable, we may in principle use the standing wave to drive electric-dipole-forbidden transitions in the particle in a single-step process\cite{Knuffman.2007} rather than a multi-step (multi-photon) process\cite{Sakurai.1967}.  We may achieve this by modulating the standing-wave intensity at the resonance frequency of the desired transition.  In our experimental system, we have demonstrated this phenomenon by modulating the intensity of the optical lattice at the resonant microwave frequency of a Rydberg-Rydberg transition and driving an atomic transition that is typically forbidden.  The advantage of the ability to drive forbidden transitions in a single step rather than through a multi-photon process is that we may avoid high field strengths that induce large unwanted AC Stark shifts of the observed transition frequencies.  As a consequence, ponderomotive spectroscopy enables the study of otherwise-forbidden electronic transitions, which is very convenient for precision spectroscopy.

Furthermore, another powerful innovation afforded by ponderomotive spectroscopy is a spatial resolution orders of magnitude better than the frequency of the transition would suggest.  Electronic transitions in a particle are typically driven by applying radiation resonant with the transition frequency.  The best possible spatial resolution will be at the diffraction limit of the applied radiation, which is on the order of the wavelength corresponding to the transition frequency and, in most cases, orders of magnitude larger than the particle size.  In contrast, in ponderomotive spectroscopy, the frequency of the applied radiation is very different from the frequency of the transition being driven.  The applied radiation is a standing wave with a wavelength on the order of the particle size.  The frequency resonant with the desired transition is introduced by intensity-modulating the standing wave. In the present report, we demonstrate ponderomotive spectroscopy by driving microwave-frequency atomic transitions (typical resolution: centimetre-scale) by intensity-modulating a standing-wave optical lattice (typical resolution: micrometre-scale).  As a result, ponderomotive spectroscopy may enable advances in quantum computing\cite{Jaksch.2000, Saffman.2010}, where single-site addressability plays a central role.
We report a successful demonstration of this new spectroscopy.  We trap $^{85}$Rb atoms in a one-dimensional standing-wave optical lattice, formed by counter-propagating 1064-nm laser beams.  We drive the $\transition$ transition by sinusoidally modulating the lattice intensity at the resonant transition frequency, found to be 38.76861(1)~GHz.  The $\transition$ transition has been chosen because it is forbidden under electric-dipole selection rules and because we expect a particularly high population transfer rate in our intensity-modulated lattice\cite{Knuffman.2007}.

\spacing{1}
\begin{figure}
\centering
\includegraphics[width=3in]{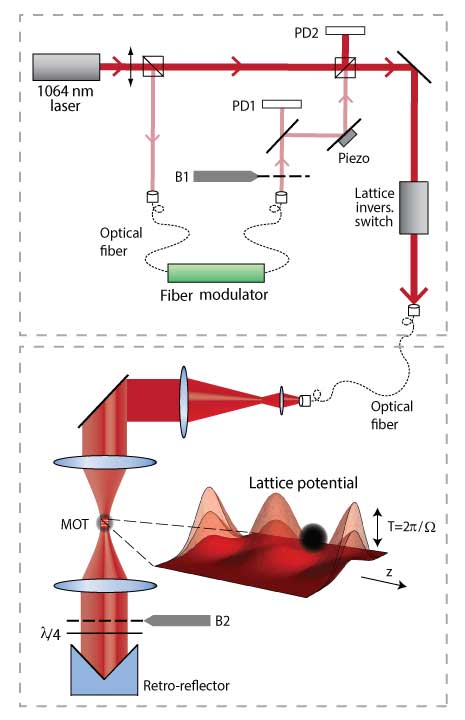}
\caption{\bf Experimental set-up.  \rm A Mach-Zehnder interferometer (top) combines two 1064-nm c.w. beams.  One beam is intensity-modulated at frequency $\Omega$ by a fibre-based electro-optic modulator (`Fibre modulator'), the operating point of which is set using photo-detector `PD1'.  Using photo-detector `PD2' and a piezo-electric transducer (`Piezo'), the interferometer is locked such that the beams add up in phase.  Rydberg atoms are laser-excited and optically trapped by lattice inversion.  The optical lattice (bottom) is formed by retro-reflecting and focusing the 1064-nm laser beam into the magneto-optical trap (`MOT').  The lattice potential is intensity-modulated with period $T$.  Beam blocks (`B1',`B2') are used for signal interpretation.  Details in Methods.}\label{figure1}
\end{figure}
\spacing{1.5}

\subsection{}
\subsection{Pulsating the optical lattice at a microwave frequency.}
To motivate the use of an intensity-modulated optical lattice for a demonstration of ponderomotive spectroscopy, we briefly examine the physics underlying the interaction between the Rydberg atom and the lattice light.  The interaction between the atom and a light field is described by the following interaction Hamiltonian\cite{Sakurai.1967}:
\begin{equation}
V_\mathrm{int} = \frac{1}{2m_e}\left(2\left|e \right|\mathbf{A} \cdot \mathbf{p} + e^2\mathbf{A} \cdot \mathbf{A}\right), \label{eq:eq2}
\end{equation}
where $\mathbf{p}$ is the Rydberg electron's momentum operator and $\mathbf{A}$ the vector potential of the light (the laser electric field $\mathbf{E}=-\frac{\partial}{\partial t} \mathbf{A})$.  The $\mathbf{A} \cdot \mathbf{p}$ term describes most types of atom-field interactions and is commonly engaged in spectroscopy\cite{Sakurai.1967}.  Transitions driven by the $\mathbf{A} \cdot \mathbf{p}$ term follow well-established spectroscopic selection rules\cite{Bethe}.  In the case of direct application of microwave radiation to Rydberg atoms, the electric-dipole selection rules apply (in first-order perturbation theory).  In contrast, the quadratic $\mathbf{A} \cdot \mathbf{A}$ (ponderomotive) term allows us to drive transitions far beyond the electric dipole selection rules in first order by providing a substantial spatial variation of the field intensity within the volume of the atom and by modulating the intensity in time at the transition frequency between the coupled states.  This motivates us to drive atomic transitions using an optical lattice that is spatially modulated with a period on the same scale as the Rydberg-atom diameter and that is also intensity-modulated at the atomic transition frequency.  In earlier work\cite{Knuffman.2007}, we proposed that, utilizing the $\mathbf{A} \cdot \mathbf{A}$ term, one can drive a wide variety of transitions beyond the usual spectroscopic selection rules.

In Figure~\ref{figure1} we show a schematic of the experimental set-up.  A continuous-wave (c.w.) 1064-nm laser beam is split in a Mach-Zehnder interferometer (top) into a low-power and high-power beam.  The low-power beam is sinusoidally modulated via an electro-optic fibre modulator driven by a tunable microwave-frequency voltage signal.  This intensity-modulated low-power beam is coherently re-combined with the unmodulated high-power beam at the exit of the interferometer.  In the atom-field interaction region (bottom), the intensity ratio between the modulated and unmodulated portions of the incident lattice beam is about 1:100.  We form the standing-wave optical lattice by retro-reflecting the lattice beam.  Cold $^{85}$Rb Rydberg atoms are trapped with centre-of-mass positions near intensity minima of the lattice\cite{Anderson.2011} (Methods).  At these locations, the intensity modulation of the lattice results, via the $\mathbf{A} \cdot \mathbf{A}$ term, in a time-periodic atom-field interaction with a leading quadratic dependence on position (needed to drive an $S \rightarrow S$ transition).  The proper combination of temporal and spatial intensity modulation is essential for utilizing the $\mathbf{A} \cdot \mathbf{A}$ term to realize the type of spectroscopy introduced in this report.

Due to the coherent mixing of the low-power, modulated lattice beam with a high-power, unmodulated beam, the time-averaged lattice depth is large enough that most atoms remain trapped in the lattice while the $\transition$ transition is probed.  The coherent mixing of the two beams is also beneficial because it enhances the modulation in the atom-field interaction region, resulting in a much larger $\transition$ coupling than would be possible with the weak modulated beam alone.  The Rabi frequency for a transition $\left|n,l,m\right\rangle \rightarrow\left|n',l',m'\right\rangle$ between two Rydberg states that are resonantly coupled by the intensity-modulated lattice is
\begin{equation}
\chi \approx \sqrt{\varepsilon} \frac{e^2}{\hbar m_e c \epsilon_0 \omega^2} \Imodamp \mathrm{J}_1 \left(\frac{\pi V_{\mathrm{IM}}}{V_\pi} \right) \left[1+\sqrt{\frac{2 \Idcamp}{\Imodamp}}\right] D_{n,l,m}^{n',l',m'},\label{eq:eq1}
\end{equation}

\noindent where $V_{\mathrm{IM}}$ is the amplitude of the microwave voltage signal that drives the fibre modulator, $V_\pi$ is the voltage difference between minimum and maximum intensity transmission through the modulator (a fixed modulator property), $\varepsilon$ is the intensity ratio at the atom location between the return and incident beams forming the lattice, $e$ is electron charge, $m_\mathrm{e}$ is electron mass, $\omega$ is the angular frequency of the optical-lattice light, and $\Imodamp$ and $\Idcamp$ are the incident intensities of the modulated and unmodulated lattice beams at the atom location, respectively.  The transition matrix element $D_{nlm}^{n'l'm'}$ (unitless) has been derived in previous work\cite{Knuffman.2007} and is 0.215 for the $\transition$ transition for an atom located at a minimum in our lattice.   This value is large compared with those of other possible transitions due to a favorable ratio between atom size and lattice period (which is on the order of one).  A detailed derivation of equation~\eqref{eq:eq1} is given in Supplementary Information.  The in-phase addition of the fields corresponding to intensities $\Imodamp$ and $\Idcamp$, using the Mach-Zehnder beam combination set-up, leads to the enhancement term in square brackets in equation~\eqref{eq:eq1}.  In our experiment, the Rabi frequency is enhanced by a factor of $\approx 20$, which aids significantly in observing the transition.  While in our experiment the enhancement afforded by the interferometric set-up is critical for a successful demonstration, different laser wavelengths, modulation schemes, sub-Doppler and evaporative cooling techniques or the use of separate trapping and modulation beams could make the interferometric set-up unnecessary.

\subsection{}
\section*{Results}

\spacing{1}
\begin{figure}
\centering
\includegraphics[width=5in]{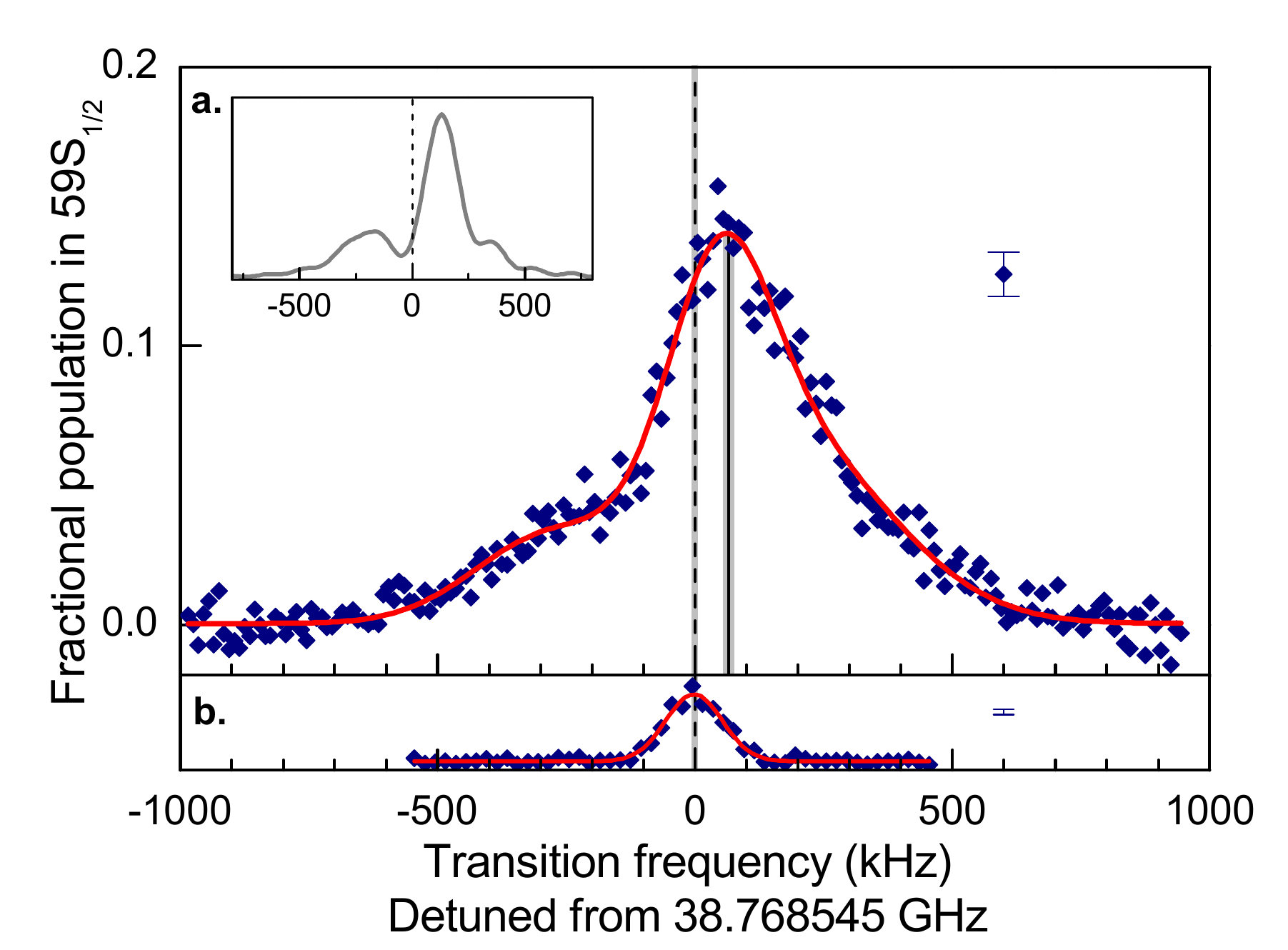}
\caption{\bf The $\mathbf{\transition}$ transition driven via intensity modulation of an optical lattice.  \rm  The fraction of atoms in $59S_{1/2}$ is shown.  Spectral-line centres and background offset are determined by Gaussian fits (red curves).  Gray areas represent s.e.m. of fit results for line centres.  \textbf{a}, Spectral line obtained by scanning the modulation frequency of the intensity-modulated optical lattice.  The data are averages of 28 scans with 200 measurements each, taken under similar experimental conditions.  Vertical error bar, s.e.m. of the 28 scans.  Line centre is at 38.76861(1)~GHz (black solid line).  Inset, simulation results.  \textbf{b}, Reference scan performed by conventional two-photon microwave spectroscopy without the optical lattice, plotted versus twice the microwave frequency.  The data are averages of 11 scans with 200 measurements each.  Line centre is at 38.768545(5)~GHz (black dashed line).}\label{figure2}
\end{figure}
\spacing{1.5}

\subsection{Spectroscopy results.}
In Figure~\ref{figure2}a we show the atomic transition $\transition$ driven via intensity-modulation of the optical lattice.  The $\transition$ spectrum is obtained by scanning the frequency of the microwave source driving the fibre modulator across the expected transition frequency.  The spectral line shown in Figure~\ref{figure2}a is proof that the $\transition$ transition has been driven in a single step (in first order), at the fundamental transition frequency.  The inset in Figure~\ref{figure2}a shows a simulated spectrum (details given in Supplementary Information).  The simulation indicates that the sub-structure in the spectral line originates from different types of centre-of-mass trajectories of the atoms in the optical lattice.  The dominant central peak results mainly from trapped atoms, whereas the small side structures result from atoms that traverse several lattice wells or that remain nearly stationary close to a lattice maximum during atom-field interaction\cite{YoungeNov.2010}.  The sub-structures, which are observed in the simulation, are not as clearly resolved in the experiment.  This is most likely due to inhomogeneous broadening.  The simulation motivates us to fit the spectrum in Figure~\ref{figure2}a as a triple-Gaussian, which yields a central-peak location of 38.76861(1)~GHz.

For reference, we have also driven the $\transition$ transition using direct application of microwave radiation at half the transition frequency; in this case the transition results from a two-step (second-order) electric-dipole coupling through the $58P_{3/2}$ off-resonant intermediate state.  In Figure~\ref{figure2}b we show the lattice-free $\transition$ spectral line, driven as a two-photon electric-dipole transition at half the transition frequency, using microwaves from a horn directed at the atom-field interaction region.  This reference measurement yields a $\transition$ transition frequency of 38.768545(5)~GHz.  The transition frequencies measured in Figures~\ref{figure2}a~and~2b are in good agreement with each other.  The slight blue-detuning of the central peak in Figure~\ref{figure2}a relative to the peak in Figure~\ref{figure2}b is due to a light shift from the optical lattice.  Calculations based on published quantum defect values\cite{Mack.2011} predict a transition frequency at 38.7686(1)~GHz, which is in agreement with the experimental measurements.  We note that for the two-photon transition shown in Figure~\ref{figure2}b the microwave-induced AC Stark shift of the transition frequency is unusually small, due to near-cancellation of the upper- and lower-level AC shifts.  We have estimated this shift of the transition frequency to be less than 500~Hz, which is insignificant in the above comparisons.

\spacing{1}
\begin{figure}
\centering
\includegraphics[width=5in]{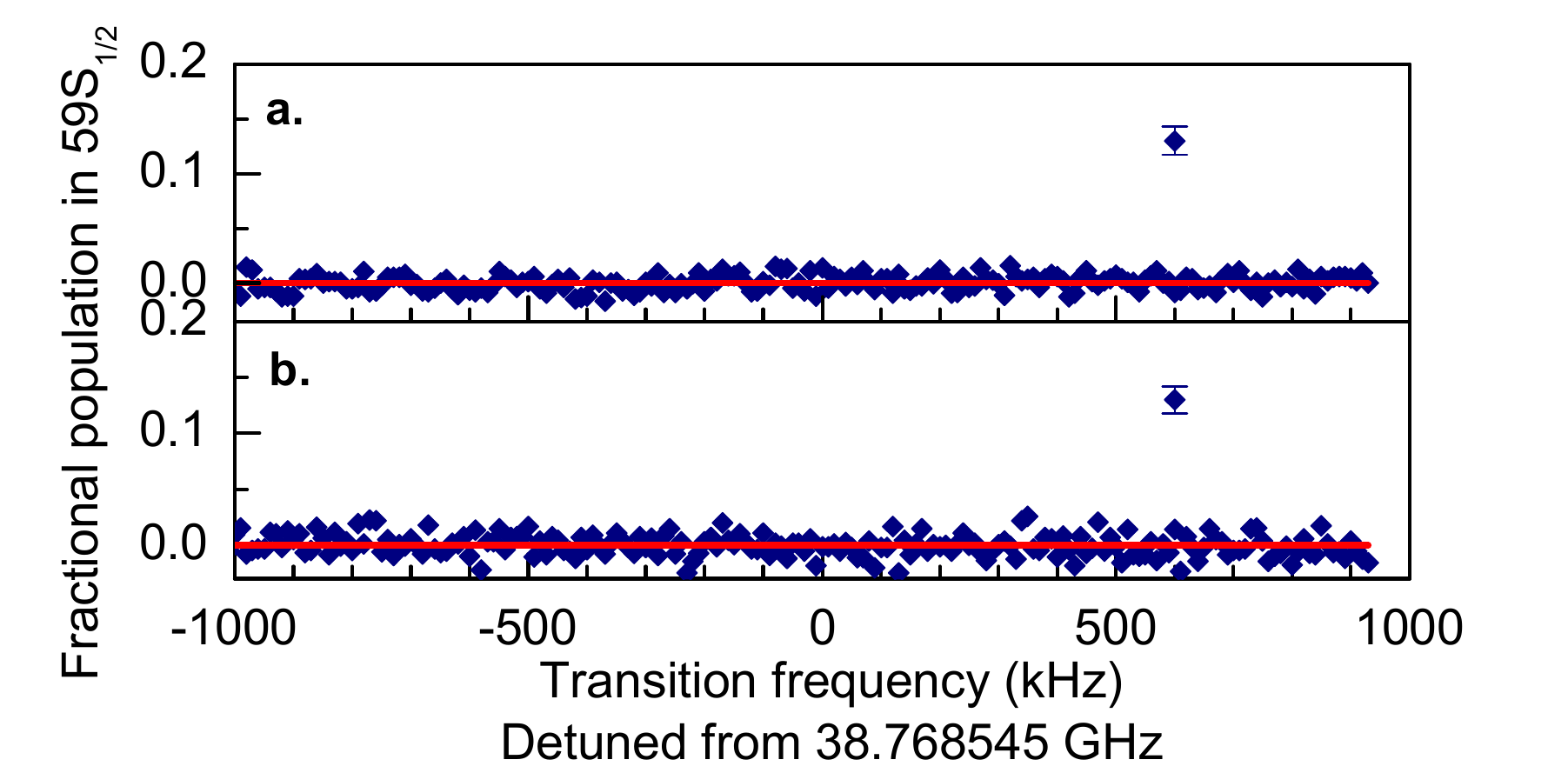}
\caption{\bf Verification of the nature of the atom-field interaction mechanism.  \rm   The fraction of atoms in $59S_{1/2}$ is recorded as the modulation frequency is scanned.  The absence of signals is important because it verifies the model presented in the text.  Background offset is determined by a constant fit (red line). \textbf{a}, Test to exclude dipole transitions caused by microwave leakage into the atom-field interaction region, performed using beam block `B1' (Figure~\ref{figure1}) during frequency scans, while leaving the microwave source at full power.  The data are averages of 16 scans with 200 measurements each.  Error bar, s.e.m. of the 16 scans.  \textbf{b}, Test to exclude a stimulated electric-dipole Raman transition, performed using beam block `B2' (Figure~\ref{figure1}) during frequency scans.  The data are averages of 10 scans with 200 measurements each.  Error bar, s.e.m. of the 10 scans.}\label{figure3}
\end{figure}
\spacing{1.5}

\subsection{}
\subsection{Testing the nature of the atom-field interaction mechanism.}
In Figure~\ref{figure3} we show two tests that prove the spectral line shown in Figure~\ref{figure2}a is indeed caused by a perturbation due to the $\mathbf{A} \cdot \mathbf{A}$ atom-field interaction term in equation~\eqref{eq:eq2}, which drives transitions via an intensity-modulated optical lattice.  First, in Figure~\ref{figure3}a we verify that the transition does not originate from a combination of microwave leakage and stray DC fields.  Although we carefully zero DC electric fields, a stray DC electric field in the atom-field interaction volume could, in principle, weakly perturb the $58S_{1/2}$ and $59S_{1/2}$ levels by adding some $P$-admixture to these levels.  Any microwave-radiation leakage into the atom-field interaction region under the simultaneous presence of a DC electric field would drive the transition as an electric-dipole transition between the weakly-perturbed $58S_{1/2}$ and $59S_{1/2}$ levels.  In order to verify that our spectral line is not due to such a coincidence, we look for a spectral line while leaving all microwave equipment fully powered but blocking the weak, intensity-modulated optical beam in the Mach-Zehnder interferometer (beam block `B1' in Figure~\ref{figure1}).  In this case, the atoms are still trapped via the high-power, unmodulated beam.  A stray DC electric field and leakage of microwave radiation into the chamber would drive the transition in a single step, as described.  In Figure~\ref{figure3}a, we scan the microwave frequency in a manner identical with the procedure used for Figure~\ref{figure2}a.  No spectral line is evident in Figure~\ref{figure3}a.  This establishes that lattice light modulation, rather than microwave leakage in the presence of a stray DC electric field, drives the observed transition.

Next, we aim to distinguish between the possible $\mathbf{A} \cdot \mathbf{p}$ and $\mathbf{A} \cdot \mathbf{A}$ transition mechanisms in equation~\eqref{eq:eq2}.  As detailed in the next paragraph, either mechanism may, in principle, drive the observed $\transition$ transition via lattice modulation.  However, these fundamentally distinct mechanisms differ in ways that can be tested experimentally.

The modulated light field contains the frequency $\omega$ and frequency sidebands $\omega \pm \Omega$, where $\omega$ is the light frequency, and $\Omega$ is the microwave modulation frequency of the light intensity (which is resonant with the $\transition$ transition).  The presence of multiple frequencies allows, in principle, for the $\mathbf{A} \cdot \mathbf{p}$ term to couple $58S_{1/2}$ to $59S_{1/2}$ in a two-step (second-order) process via a stimulated Raman transition through one or more states that have an energy separation $\approx \hbar \omega$ from the $58S_{1/2}$ and $59S_{1/2}$ levels.  Following dipole selection rules, this mechanism would involve optical $S \rightarrow P$ and $P \rightarrow S$ electric-dipole transitions through distant intermediate $P$-states.  The Raman coupling mechanism would be effective in both running-wave or standing-wave laser fields.  On the other hand, the coupling due to the $\mathbf{A} \cdot \mathbf{A}$ term is proportional to $\left\langle 59S_{1/2} \right| I \left(z\right) \left|58S_{1/2} \right\rangle \sin \left(\Omega t\right)$, where $I \left(z\right)$ is the light intensity (Figure~\ref{figure1}). For this coupling to be effective in first order, two conditions must be simultaneously fulfilled:  the light intensity must substantially vary as a function of position $z$ within the volume of the atom, and the modulation frequency $\Omega$ must correspond to the energy-level difference.  Without spatial variation of $I \left(z\right)$, the coupling vanishes due to the orthogonality of the atomic states.  Hence, the $\mathbf{A} \cdot \mathbf{A}$ coupling can be turned on or off by providing an intensity-modulated standing wave or a running wave, respectively (where the running-wave beam may or may not be intensity-modulated in time).

To test this, in Figure~\ref{figure3}b we exchange the intensity-modulated standing-wave optical lattice for an intensity-modulated running-wave beam by blocking the retro-reflected lattice beam with beam block `B2' in Figure~\ref{figure1}.  We then scan the microwave frequency in a manner identical with the procedure used for Figure~\ref{figure2}a.  No spectral line is evident in Figure~\ref{figure3}b.  Therefore, the transition mechanism responsible for the spectral line observed in Figure~\ref{figure2}a is indeed a single-step atom-field interaction arising from the modulated optical standing-wave intensity (via a first-order $\mathbf{A} \cdot \mathbf{A}$ interaction), not a two-step electric-dipole Raman coupling process arising from the frequency sidebands in the light field (via a second-order $\mathbf{A} \cdot \mathbf{p}$ interaction).

\spacing{1}
\begin{figure}
\centering
\includegraphics[width=3.5in]{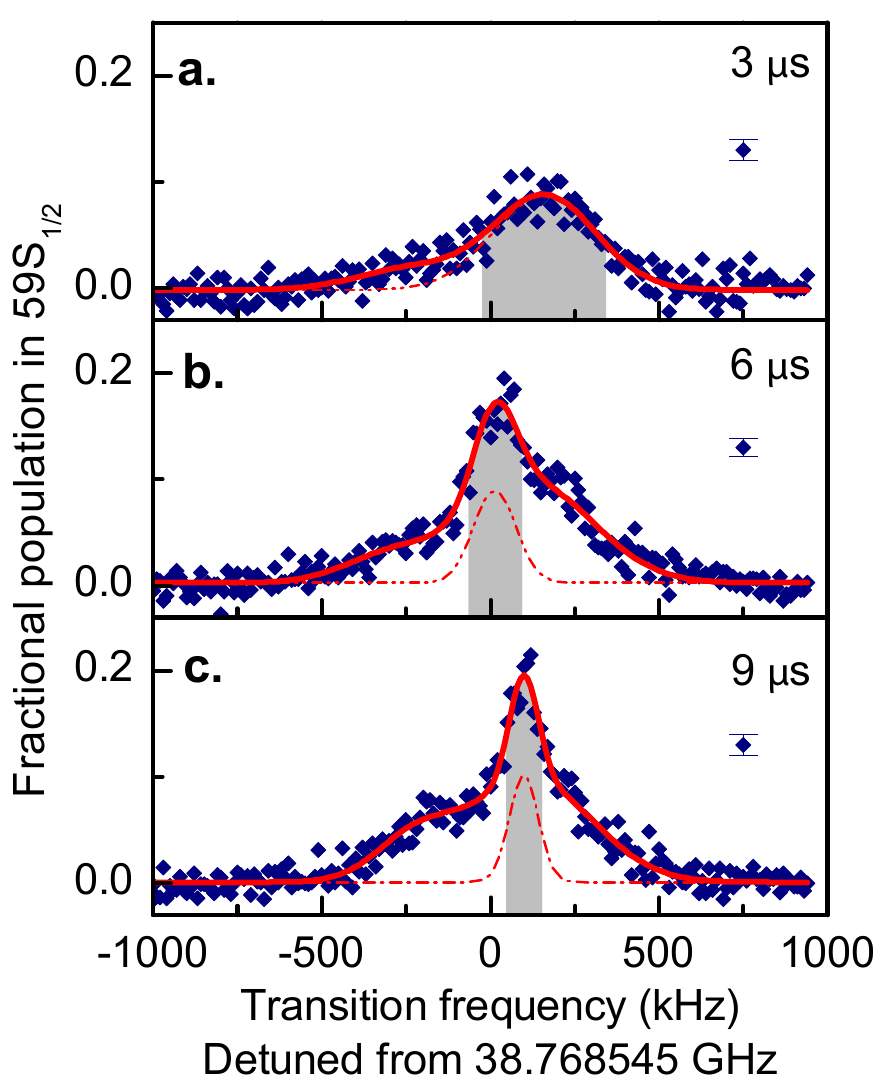}
\caption{\bf Dependence of intensity-modulation-driven $\mathbf{\transition}$ spectral line on interaction time $t_{\mathrm{int}}$ with the modulated lattice.  \rm \textbf{a-c}, For each $t_{\mathrm{int}}$, the fraction of atoms in $59S_{1/2}$ is recorded while scanning the microwave modulation frequency.  The data are averages of 8 scans with 200 measurements each.  Error bar, s.e.m. of the 8 scans. The data for $t_{\mathrm{int}}=$~3 (6 and 9)~$\mu$s are fit with double- (triple-) peak Gaussians (red solid curves).  The FWHM (gray areas) of the dominant Gaussians (red dashed curves) decrease with increasing $t_{\mathrm{int}}$, and the heights increase.}\label{figure4}
\end{figure}
\spacing{1.5}

\subsection{}
\subsection{Dependence on experimental parameters.}
Here we characterize the dependence of ponderomotive spectroscopy on several experimental parameters.  In Figure~\ref{figure4} we show the scaling behaviour of the spectral line width and amplitude on the atom-field interaction time.  Experimentally, the interaction time $t_{\mathrm{int}}$ is defined as the time between when the atoms are excited to the $58S_{1/2}$ state and when the atoms are ionized for detection (Methods), typically $t_{\mathrm{int}} = 6~\mu$s.  During the atom-field interaction time, the transitions are driven by the intensity-modulated lattice (which is always on), and the atoms undergo a square-pulse coupling to state $59S_{1/2}$ of duration $t_{\mathrm{int}}$.  In the limit of weak saturation and assuming a Fourier-limited spectral profile of the driving field, the full-width-at-half maximum (FWHM) of the spectral line is expected to decrease with increased interaction time as $\approx 0.9 / t_{\mathrm{int}}$.  This agrees with the trend observed in Figure~\ref{figure4}.  A double-Gaussian fit of the 3~$\mu$s spectral line and triple-Gaussian fits of the 6-and-9-$\mu$s spectral lines (red solid curves in Figure~\ref{figure4}) indicate that the FWHM of the dominant Gaussian components in each line (red dashed curves) are within 20\% of the Fourier limit.  

Examining the spectral lines in Figure~\ref{figure4} further, we also find that the maximum signal height approximately doubles between 3~and~6~$\mu$s; the additional increase between 6~and~9~$\mu$s is relatively minor.  This observation is consistent with a Rabi frequency within the range of 50-100~kHz.  This result is in qualitative agreement with our calculated Rabi frequency presented in Supplementary Information.

\spacing{1}
\begin{figure}
\centering
\includegraphics[width=3.5in]{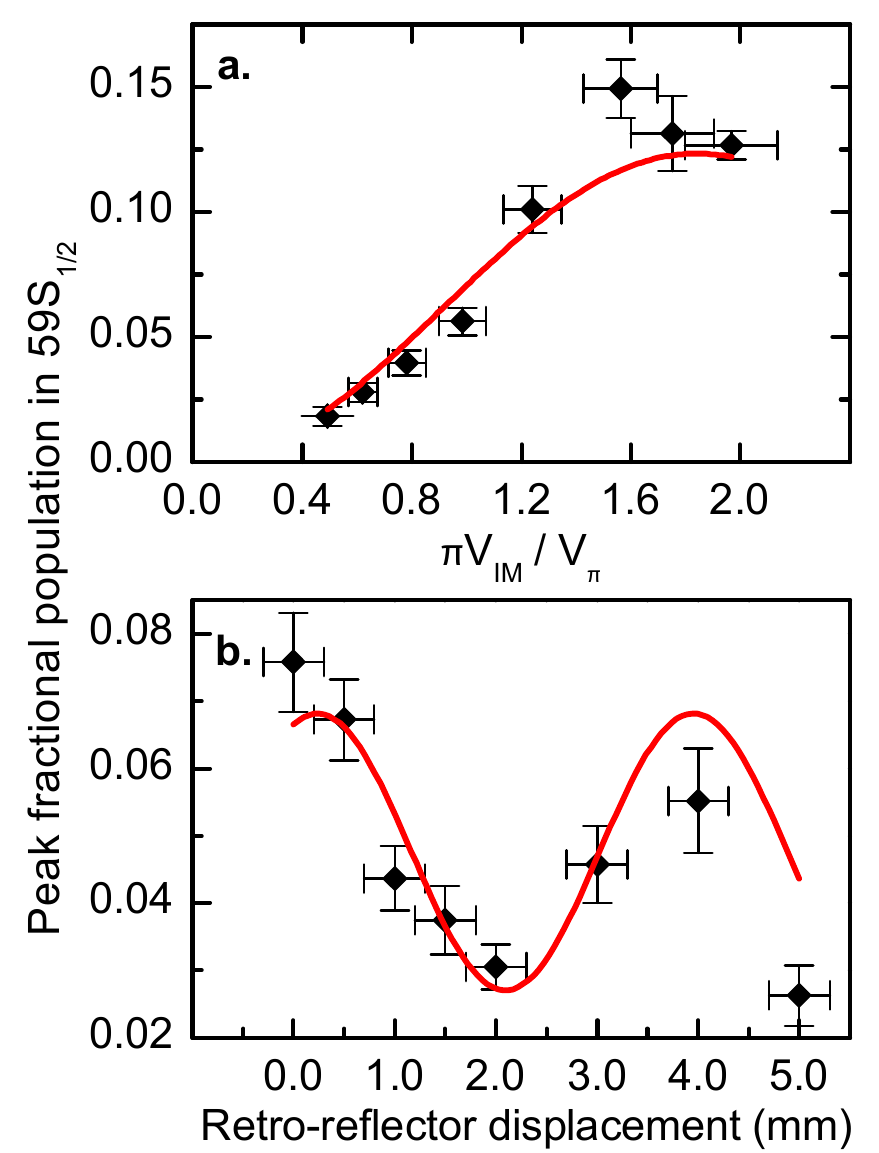}
\caption{\bf Dependence of intensity-modulation-driven $\mathbf{\transition}$ spectral-line height on experimental parameters.  \rm The peak fraction of $59S_{1/2}$ atoms is recorded while varying experimental parameters.  Each data point represents the observed peak height for an average of 5 (\textbf{a}) or 3 (\textbf{b}) scans with 200 measurements each.  Vertical error bars, uncertainty of peak height.  \textbf{a}, Dependence on $\pi V_{\mathrm{IM}} / V_\pi$, where we vary the microwave voltage amplitude $V_{\mathrm{IM}}$ applied to the fibre modulator ($V_\pi$ is fixed).  Horizontal error bars are due to the uncertainty in $V_\pi$.  Red curve, fit proportional to $\mathrm{J}_1^2\left(\pi V_{\mathrm{IM}} / V_\pi\right)$.  \textbf{b}, Dependence on lattice retro-reflector distance from the atoms.  Horizontal error bars, distance uncertainty.  A sinusoidal curve (red) with a period of 4~mm is plotted for reference.} \label{figure5}
\end{figure}
\spacing{1.5}

In Figure~\ref{figure5} we summarize the dependence of the spectral-line height on additional experimental parameters.  In Figure~\ref{figure5}a we investigate the dependence of spectral-line height on modulation strength, which is controlled by varying the amplitude of the microwave voltage signal $V_{\mathrm{IM}}$ that drives the fibre modulator.  As can be seen in equation~\eqref{eq:eq1}, the Rabi frequency $\chi$ has a first-order Bessel function ($\mathrm{J}_1$) dependence on $V_{\mathrm{IM}}$.  Because the height of the spectral line indicates the peak fraction of population in $59S_{1/2}$, we expect the height to scale as $\left(\chi t_{\mathrm{int}}\right)^2 \propto \mathrm{J}_1^2\left(\pi V_{\mathrm{IM}} /V_\pi\right)$, for fixed $t_{\mathrm{int}}$ and in the limit of weak saturation.  In Figure~\ref{figure5}a, we plot the spectral line height as a function of the Bessel function argument, $\pi V_{\mathrm{IM}} /V_\pi$, as we vary $V_{\mathrm{IM}}$.  A  $\mathrm{J}_1^2\left(\pi V_{\mathrm{IM}} /V_\pi\right)$ fit to the data yields good agreement.

The spectral-line height also depends on the distance of the retro-reflector from the atoms.  Both the incident and the retro-reflected intensity-modulated lattice beams can be viewed as periodic sequences of pulses with a repetition frequency $\Omega$.  The spectral-line height is maximal if the pulse trains of the incident and retro-reflected lattice beams arrive synchronously at the atoms' location.  Considering the time delay between the retro-reflected and incident pulses, it is seen that the spectral-line height should sinusoidally vary with the position of the retro-reflector mirror with a period of 4~mm.  In Figure~\ref{figure5}b we plot the spectral-line height as a function of retro-reflector displacement over a range of a few millimetres.  A sinusoidal curve with a period of 4~mm has also been plotted for reference.  We observe good qualitative agreement, with deviations attributed to alignment drift of the excitation beams during acquisition of multiple data scans.  This test provides another verification of our interpretation of the transition mechanism.

In conclusion, we have successfully demonstrated ponderomotive spectroscopy for the first time.  We have demonstrated major advantages over standard spectroscopy, including improved spatial addressability and flexible transition rules, which are relevant in a broad range of applications.  Using a temporally- and spatially-modulated ponderomotive interaction, we have driven an atomic microwave transition forbidden by established electric-dipole selection rules, with a spatial resolution in the micrometre range.  In our specific case, we have demonstrated ponderomotive spectroscopy using cold Rydberg atoms and an intensity-modulated standing-wave laser beam.  One immediate application of this demonstration is in quantum computing\cite{Saffman.2010}$^,$\cite{Weimer.2010}, where single-site addressability plays a central role.  Another application is in precision measurement of atomic characteristics\cite{Han.2006} and physical constants (e.g. the Rydberg constant\cite{Kleppner.1997}, leading to the proton size\cite{Pohl.2010}); there, flexible spectroscopic transition rules will be very convenient.  In the future, one may also explore the possibility of extending ponderomotive spectroscopy to smaller-sized atoms or molecules trapped in shorter-wavelength optical lattices.

\newpage

\begin{methods}

\subsection{Preparation of cold $^{85}$Rb Rydberg atoms in an optical lattice.}
Initially, $^{85}$Rb atoms are cooled to a temperature of about 150 $\mu$K in a magneto-optical trap (MOT).  A one-dimensional, 1064-nm optical lattice is applied to the MOT region.  The optical lattice is formed by focusing an incident laser beam into the MOT region, retro-reflecting and re-focusing it.  The incident beam interferes with the return beam and forms a standing wave.  Pointing stability is greatly improved by using a retro-reflector rather than a plane mirror.  See Figure~\ref{figure1} in the main text for an illustration.  The lattice has 880~mW of incident power at the MOT region with an 11-micrometre beam waist radius, and 490~mW of return power with an approximately 37-micrometre beam waist radius.  See Supplementary Information for a discussion of these numbers.  Both MOT magnetic field and lattice light remain on throughout a data scan, while the MOT light is turned off during Rydberg atom excitation and probing.

The Rydberg state is excited via a two-stage excitation $5S_{1/2} \rightarrow 5P_{3/2} \rightarrow 58S_{1/2}$ using 780-nm and 480-nm lasers, detuned from the intermediate $5P_{3/2}$ level by $\approx$~1~GHz.  Immediately after Rydberg-atom excitation, the lattice is inverted using an electro-optic component, required to efficiently trap Rydberg atoms at intensity minima.  The Rydberg atoms are trapped in the optical field using the energy shift of the quasi-free Rydberg electron\cite{Anderson.2011}.  During a data scan, the Rydberg atom excitation laser wavelength is tuned so as to produce Rydberg atoms in the optical lattice at an average of 0.5 Rydberg atoms per experimental cycle.  This ensures a negligible probability of atom-atom interactions.

\subsection{Zeroing the fields.}
The electric field is zeroed prior to a series of data runs by performing Stark spectroscopy\cite{Frey.1993} of the $58D_{3/2}$ and $58D_{5/2}$ fine-structure levels, while scanning the potentials applied to a set of electric-field compensation electrodes.  The residual field is less than 60~mV/cm.  The MOT magnetic field, which is on during the experiment, does not cause Zeeman shifts of the $\transition$ transition.

\subsection{Lattice modulation.}
While $58S_{1/2}$ Rydberg atoms are trapped in optical lattice wells, the intensity of the lattice is modulated at the expected $\transition$ transition frequency.  The electro-optic modulator used to perform the modulation is a fibre-coupled, polarization-maintaining, z-cut lithium niobate modulator, tunable from DC to 40~GHz.  The modulator has an input optical power limit of 200~mW, which is, in the present experiment, insufficient for Rydberg-atom trapping.  In a Mach-Zehnder-type interferometric set-up, we split the 1064-nm laser into a high-power (3.9~W), unmodulated beam and a low-power (190~mW), intensity-modulated beam, which is passed through the fibre modulator.  The two beams are coherently re-combined at the exit beam-splitter of the Mach-Zehnder interferometer.  The re-combined beam incident at the MOT region has approximately 1-W of average power, which is sufficient for Rydberg atom trapping.  See Supplementary Information.

\subsection{Operating point of the fibre modulator.}
The fibre modulator has two voltage inputs, one for the microwave voltage signal and another for a DC bias voltage.  As described by equations (1) and (2) in Supplementary Information, the intensity transmitted through the fibre modulator depends on the values of these voltage inputs.  For most of the work, the amplitude of the microwave signal is set to $V_{\mathrm{IM}} = V_\pi / 2$, which yields maximal intensity modulation when the DC bias voltage is set for a time-averaged transmission through the modulator that equals 50\% of the maximum possible.  Due to thermal drifts in the fibre modulator, the DC bias voltage must be actively regulated to maintain this operating point.  The lock circuit utilizes a photo-detector (`PD1' in Figure~\ref{figure1}) and a PID regulator.

\subsection{Locking the interferometer.}
Due to drifts in the optical path length difference between the arms of the Mach-Zehnder interferometer, the path length of one of the interferometer arms must be actively regulated to maintain a fixed phase difference at the recombination beam-splitter.  This phase difference is locked so that the intensity sent to the experiment is at a maximum (by maintaining an intensity minimum at the unused output of the recombination beam-splitter).  The lock circuit utilizes a photo-detector (`PD2' in Figure~\ref{figure1}), a mirror mounted on a piezo-electric transducer (`Piezo' in Figure~\ref{figure1}), and a PID regulator.

\subsection{Detection.}
The spectral line is detected through state-selective field ionization\cite{Gallagher.1994}, in which Rydberg atoms are ionized by a ramped electric field.  Freed electrons are detected by a micro-channel plate, and detections on the micro-channel plate are registered by a pulse counter.  Counting gates are synchronized with the field ionization ramp to enable state-selective detection of the $58S_{1/2}$ and $59S_{1/2}$ Rydberg levels.

\subsection{Read-out protocol.}
During a data scan, the microwave frequency of the intensity-modulation is stepped across the expected $\transition$ resonance frequency.  At each microwave frequency step, the pulse counter registers counts for 200 experimental cycles.  Average counts per cycle are recorded before advancing to the next frequency step.  The interferometer lock status is queried before and after each set of 200 experimental cycles.  If either query indicates an unlocked interferometer, the data for that frequency step is ignored and re-taken.
\end{methods}

%% Here is the endmatter stuff: Supplementary Info, etc.
%% Use \item's to separate, default label is "Acknowledgements"

\begin{addendum}
 \item[Acknowledgements] S.E.A. acknowledges support from DOE SCGF.  This work was supported by NSF Grant No. PHY-1205559 and NIST Grant No. 60NANB12D268.
 \item[Author Contributions] All authors contributed extensively to the work presented in this paper.
 \item[Author Information] Reprints and permissions information is available at www.nature.com/ncomms. The authors declare no competing financial interests. Correspondence and requests for materials should be addressed to K.R.M.~(kaimoore@umich.edu).
\end{addendum}

\newpage
\setcounter{page}{1}

\section*{Supplementary Information}

\section{Derivation of the Rabi frequency}
\subsection{Intensity of the modulated running wave.}

The optical power ($\lambda$~=~1064~nm) transmitted through the fibre electro-optic modulator is

\begin{equation}
P_\mathrm{T} = P_{\mathrm{low}}\sin^2 \left(\frac{\pi}{2} \frac{V_{\mathrm{total}}}{V_\pi}\right), \label{eq:eq1}
\end{equation}

\noindent where $P_{\mathrm{low}}$ is the maximum possible output of the fibre modulator, and $V_\pi$ is the voltage difference between adjacent power transmission maxima and minima (a fixed property of the fibre modulator).  $V_{\mathrm{total}}$, the total voltage on the fibre modulator, is

\begin{equation}
V_{\mathrm{total}} = V_{\mathrm{DC}}\left( t\right) + V_\mathrm{IM} \sin \left(\Omega t \right) + V_{\mathrm{drift}}\left( t \right),\label{eq:eq2}
\end{equation}

\noindent where $V_{\mathrm{DC}}\left( t\right)$ is the applied DC bias voltage and $V_\mathrm{IM}$ is the amplitude of the intensity-modulating microwave voltage signal with frequency $\Omega$.  Bias drifts due to thermal and environmental effects are accounted for via a slowly-varying effective drift voltage $V_{\mathrm{drift}}\left( t\right)$.

Experimentally, $V_{\mathrm{DC}}\left( t\right)$ is actively regulated such that $V_{\mathrm{DC}}\left( t\right) + V_{\mathrm{drift}}\left( t\right)$ is locked to the inflection point of the transmission curve given by supplementary equation~\eqref{eq:eq1}.  With the fibre modulator locked, the power transmission curve is

\begin{equation}
P_\mathrm{T} = P_{\mathrm{low}}\left( \frac{1}{2} + \frac{1}{2}\sin \left(\frac{\pi V_{\mathrm{IM}}}{V_\pi} \sin \left( \Omega t \right) \right) \right).\label{eq:eq3}
\end{equation}

\noindent Most experiments are performed at full modulation, $V_{\mathrm{IM}} = V_\pi / 2$.

At the location of the atoms, the intensity is determined by the incident (1/e$^2$) beam waist radius $w_{\mathrm{inc}}$~=~11~$\mu$m and the incident beam power, which is measured (in the absence of modulation) to be $\Pmodamp=$5-10~mW.  (This value is much less than $P_{\mathrm{low}}$ in supplementary equation~\eqref{eq:eq1} due to inefficiencies in the optical system.)  Therefore, the maximum intensity of the modulated running-wave beam incident at the location of the atoms is
\begin{equation}
\Imodamp = \frac{2 \Pmodamp}{\pi w_{\mathrm{inc}}^2}.\label{eq:eq4}
\end{equation}

\noindent The intensity of the modulated (low-power) running-wave beam incident at the location of the atoms can be determined by supplementary equations~\eqref{eq:eq3} and \eqref{eq:eq4}, along with a Jacobi-Anger substitution, and is

\begin{equation}
I_{\mathrm{rw}}^{\mathrm{mod}} = \Imodamp \left(\frac{1}{2} + \sum_{q=1}^{\infty} \mathrm{J}_{\mathrm{2q-1}} \left(\frac{\pi V_{\mathrm{IM}}}{V_\pi} \right) \sin \left( \left(2q-1\right) \Omega t\right) \right).\label{eq:eq5}
\end{equation}

\subsection{Intensity expression for modulated standing wave.}

The power transmission efficiency of the optics in the lattice return beam path has been determined to be 56\%.  Optical aberrations and a displacement of the retro-reflector position from the optical axis have been estimated to yield an effective return beam waist radius of $w_{\mathrm{ret}}$~=~37~$\mu$m (also see section 2 of supplementary equations).  Therefore, the ratio of the return and incident intensity on the optical axis is $\varepsilon$~=~9\%.  After adding the fields of the incident and return beams coherently, the intensity of the modulated (low-power) standing-wave beam is

\begin{align}
I_{\mathrm{sw}}^{\mathrm{mod}} &= \Imodamp \left(\frac{1}{2} + \sum_{q=1}^{\infty} \mathrm{J}_{\mathrm{2q-1}} \left(\frac{\pi V_{\mathrm{IM}}}{V_\pi} \right) \sin \left( \left( 2q-1\right) \Omega t\right) \right) \nonumber\\
&\qquad {} \times \Big( 1 + \varepsilon + 2\sqrt{\varepsilon}\cos \left(2 k z\right) \Big), \label{eq:eq6}
\end{align}

\noindent where $k$ is the optical wavenumber $k= 2\pi / \lambda$.

\subsection{Intensity expression for unmodulated standing wave.}

Similar to supplementary equation~\eqref{eq:eq4}, the intensity of the unmodulated running-wave beam at the location of the atoms is

\begin{equation}
\Idcamp = \frac{2 \Pdcamp}{\pi w_{\mathrm{inc}}^2},\label{eq:eq7}
\end{equation}

\noindent where beam power $\Pdcamp$ is measured to be 900~mW.  Therefore, at the location of the atoms, the unmodulated (high-power) beam will form a standing wave, described by an expression similar to supplementary equation~\eqref{eq:eq6}, albeit without the time-dependent envelope function.  This expression is

\begin{equation}
I_{\mathrm{sw}}^{\mathrm{unmod}} = \Idcamp \left( 1 + \varepsilon + 2\sqrt{\varepsilon}\cos \left(2 k z\right) \right). \label{eq:eq8}
\end{equation}

\subsection{Total standing-wave intensity.}

The fields of the modulated and unmodulated standing waves add up coherently to yield an intensity,

\begin{align}
I_{\mathrm{sw}}^{\mathrm{mod+unmod}} &= \Bigg[ \Imodamp \left(\frac{1}{2} + \sum_{q=1}^{\infty} \mathrm{J}_{\mathrm{2q-1}} \left(\frac{\pi V_{\mathrm{IM}}}{V_\pi} \right) \sin \left( \left( 2q-1\right) \Omega t\right) \right) \Bigg.\nonumber\\
&\qquad {} + 2 \sqrt{\Imodamp \Idcamp \left(\frac{1}{2} + \sum_{q=1}^{\infty} \mathrm{J}_{\mathrm{2q-1}} \left(\frac{\pi V_{\mathrm{IM}}}{V_\pi} \right) \sin \left( \left( 2q-1\right) \Omega t\right)\right)}\nonumber\\
&\qquad \Bigg. {} + \Idcamp \Bigg] \times \Big( 1 + \varepsilon + 2\sqrt{\varepsilon}\cos \left(2 k z\right) \Big). \label{eq:eq9}
\end{align}

\noindent This expression for the total standing-wave intensity at the location of the atoms includes any harmonics of the microwave frequency $\Omega$.

\subsection{Rabi frequency.}
We consider a two-level atomic system that has a transition resonant with the modulation frequency $\Omega$ and that is insensitive to higher harmonics (a valid consideration for this experiment).  The light intensity must also be spatially modulated within the volume of the atom for the $\mathbf{A} \cdot \mathbf{A}$ term in equation (1) of the main text to drive transitions.  After a spectral analysis of supplementary equation~\eqref{eq:eq9}, the relevant spatially- and temporally-modulated component of the standing-wave intensity is

\begin{align}
\Iomega &\approx 2\sqrt{\varepsilon} \Imodamp \mathrm{J}_1 \left(\frac{\pi V_{\mathrm{IM}}}{V_\pi} \right) \left(1+\sqrt{\frac{2 \Idcamp}{\Imodamp}}\right) \sin \left(\Omega t\right)\cos \left(2 k z\right)\nonumber\\
&=: F_\Omega \sin\left(\Omega t\right) \cos\left(2 k z\right). \label{eq:eq12}
\end{align}

\noindent  Averaged over one optical cycle, this component of the intensity leads to the term in $\mathbf{A} \cdot \mathbf{A}$ that is important in lattice modulation spectroscopy,

\begin{equation}
A_\Omega^2\left( z, t \right) = \frac{\Iomega}{c \epsilon_0 \omega^2} = \frac{F_\Omega}{c \epsilon_0 \omega^2} \sin\left(\Omega t\right) \cos\left(2 k z\right).\label{eq:eq13}
\end{equation}

\noindent It is important to note that the time-averaging needed to arrive at supplementary equation~\eqref{eq:eq13} is over one optical period ($\approx$ 4~femtoseconds), not over the lattice modulation period ($\approx$ 25~picoseconds).  In the atom-field interaction Hamiltonian (equation (1) in the main text), the corresponding term is

\begin{equation}
W_\Omega = \frac{e^2 F_\Omega}{2 m_e c \epsilon_0 \omega^2} \sin\left(\Omega t\right) \cos\left(2 k z\right) =: B\sin\left(\Omega t\right) \cos\left(2 k z\right).\label{eq:eq14}
\end{equation}

\noindent Following a previous proposal\citeKnuffman, $W_\Omega$ is the driving term in the spatially- and temporally-modulated interaction potential.  The variable $B$ in supplementary equation~\eqref{eq:eq14} is identical with the spatial/temporal modulation amplitude used in Ref.~\citeKnuffman.  As shown in Ref.~\citeKnuffman, the Rabi frequency for an atom is given by

\begin{equation}
\chi \approx \sqrt{\varepsilon} \frac{e^2}{
\hbar m_e c \epsilon_0 \omega^2} \Imodamp \mathrm{J}_1 \left(\frac{\pi V_{\mathrm{IM}}}{V_\pi} \right) \left(1+\sqrt{\frac{2 \Idcamp}{\Imodamp}}\right) D_{n,l,m}^{n',l',m'}. \label{eq:eq11}
\end{equation}

\noindent Here, $D_{n,l,m}^{n',l',m'}$ is the (unitless) transition matrix element between states $\left|n,l,m\right\rangle \rightarrow \left|n',l',m'\right\rangle$ that are resonantly coupled by frequency $\Omega$.  This matrix element includes a $\cos\left(2 k z_0\right)$ dependence, where $z_0$ is the centre-of-mass position of the Rydberg atom relative to the nearest optical-lattice intensity minimum.  For atoms located at lattice intensity minima (or maxima), the light intensity has a leading term proportional to $z^2$ and the matrix element for the transition used in the experiment, $\left\langle 59S_{1/2} \right| z^2 \left|58S_{1/2} \right\rangle$, is large.  Numerical calculations\citeKnuffmanspace show $D_{58S}^{59S}$ = 0.215 at lattice minima (and -0.215 at maxima).  Using the experimental parameters given in this supplementary material, we find a Rabi-frequency estimate of $2\pi \times 100$~kHz and a population inversion time of 5~$\mu$s.

If the $\left|59S\right\rangle$ population is small, it has a quadratic dependence on the Rabi frequency.  As described in section 2 of supplementary equations below, our atoms have velocity and position distributions.  After performing a weighted average of the population in the $\left|59S\right\rangle$ target state over these different velocity and position classes, an approximate dependence on a quadratic first-order Bessel function of the field-modulating voltage amplitude, $\mathrm{J}_1^2\left(\pi V_{\mathrm{IM}} /V_\pi\right)$, remains.  This is our motivation for exploring the population dependence on the argument of this quadratic Bessel function in Figure 5a of the main text.
\subsection{}
\subsection{}
\subsection{}
\section{Spectrum simulation}
\subsection{}
In the simulation, initial positions and velocities of $5S_{1/2}$ atom ensembles are determined from a Maxwell-Boltzmann distribution in the ground-state trapping potential.  The ground-state atom temperature, $T$, is on the order of the rubidium Doppler cooling temperature ($150~\mu$K).  The optical excitation $5S_{1/2} \rightarrow 58S_{1/2}$ is assumed to be resonant at the lattice intensity maxima, where the $5S_{1/2}$ atoms collect.  The atom-lattice interaction times are chosen consistent with the timing used in the experiment.  In the simulation the lattice is inverted upon Rydberg atom excitation (as in the experiment).

The classical centre-of-mass Rydberg-atom trajectories follow from the trapping potential calculated for the Rydberg levels\citeYounge. The quantum evolution in the internal state space $\left\{\left|58S\right\rangle,\left|59S\right\rangle \right\}$ driven by the lattice intensity modulation is computed along these trajectories by integrating the time-dependent Schr{\"o}dinger equation. Both the Rabi frequency and the detuning between modulation frequency and atomic transition frequency depend on position (which for a moving atom is time-dependent).  In the inset in Figure 2a of the main text, we show a simulated spectrum obtained for our experimental conditions.  Most parameters in the simulation have values known from the experiment.  The Rabi frequency depends on experimental parameters as specified in supplementary equation~\eqref{eq:eq11}. The measured powers of the incident and return lattice beams and the measured 1/e$^2$ radius of the incident lattice beam (11~$\mu$m) are also entered as fixed values. The only free fit parameters of the simulation are the initial atom temperature $T$ and the effective 1/e$^2$ radius of the return lattice beam, $w_{\mathrm{ret}}$, which could not be measured.  Good agreement between experimental and simulated microwave spectra is found for $w_{\mathrm{ret}}$= 37~$\mu$m and $T \approx 100$~$\mu$K, which is in line with reasonable expectations.

\subsection{}
\section*{Supplementary References}
\subsection{}
\noindent [1$^*$] Knuffman, B. \& Raithel, G. Multipole transitions of Rydberg atoms in modulated ponderomotive potentials. \it Phys. Rev. A. \rm \bf 75\rm, 053401 (2007)

\noindent [2$^*$] Younge, K.C., Anderson, S.E. \& Raithel, G. Adiabatic potentials for Rydberg atoms in a ponderomotive optical lattice. \it New J. Phys. \rm \bf 12\rm, 023031 (2010)

\end{document}